\documentclass[aps,pre,groupedaddress,twocolumn,showpacs]{revtex4-1}

\usepackage{amsmath}
\usepackage{amsfonts}
\usepackage{amssymb}

\usepackage{graphicx}
\usepackage{subfigure}

\renewcommand{\vec}[1]{{\mathbf{#1}}}

\begin{document}

\title{Convection in nanofluids with a particle-concentration-dependent thermal conductivity}

\author{Martin Gl\"assl}
\author{Markus Hilt}
\author{Walter Zimmermann}
\email[]{walter.zimmermann@uni-bayreuth.de}

\affiliation{Theoretische Physik I, Universit\"at Bayreuth, 95440 Bayreuth, Germany}

\date{September 16, 2010 (submitted to Physical Review E)}

\begin{abstract}
Thermal convection in nanofluids is investigated by means of a continuum model
for binary-fluid mixtures, with a thermal conductivity depending on the local
concentration of colloidal particles. The applied temperature difference between the
upper and the lower boundary leads via the Soret effect to a variation of the
colloid concentration and therefore to a spatially varying heat conductivity.
An increasing difference between the heat conductivity of the mixture near
the colder and the warmer boundary results in a shift of the onset of convection
to higher values of the Rayleigh number for positive values of the separation
ratio $\psi>0$ and to smaller values in the range $\psi<0$. Beyond some critical
difference of the thermal conductivity between the two boundaries, we find
an oscillatory onset of convection not only for $\psi<0$, but also within a finite range 
of $\psi>0$. 
This range can be extended by increasing the difference in the thermal conductivity 
and it is bounded by two codimension-2 bifurcations.
\end{abstract}


\maketitle

\section{Introduction\label{sec: intro}}  

Thermal convection plays a central role in geophysics \cite{Bercovici:2009,Busse:89.2}, 
atmospheric dynamics \cite{Houze:1994} and various technical applications, whereof  
nanofluids were recently identified as efficient heat-transfer substances 
\cite{Eastman:2004.1,Buongiorno:2006.1,Buongiorno:2009.1,ChoiSUS:2009.1}.
Rayleigh-B\'enard convection with its numerous variants is also a classical lab experiment
for studying generic phenomena of nonlinear dynamics and pattern formation,
which occur in this system as in several other fields of natural science
\cite{Busse:89.2,CrossHo,BOPeAh:2000.1,Cross:2009}.
In most of the related theoretical studies the Oberbeck-Boussinesq (OB)
approximation for thermal convection is used, where constant material parameters
independent of the thermodynamic variables are assumed, except the temperature-dependent
density within the buoyancy term, which is the essential driving force of convection. 

Non-Boussinesq contributions to the governing equations for convective systems
are often required to model various phenomena in an appropriate way.
Around $4^{\circ} C$, for instance, the linear term in the thermal expansion of water vanishes
and the quadratic contribution has to be taken into account. It is well known that this
modification changes the symmetry condition in a thin fluid layer heated from below,
leading to hexagonal convection patterns instead of stripe patterns close to the onset
of convection \cite{Busse:1967.1,CrossHo}. Strongly varying material properties in the
Earth's mantle are a major motivation for using a temperature-dependent viscosity in models
of thermal convection in single component fluids, which is another non-Boussinesq contribution 
\cite{Palm:1960.1,Turcotte:1971.1,Busse:85.1,Christensen:1991.1,Yuen:1995.1,OgawaM:2008.1}.
A further example is a temperature dependent thermal conductivity, being considered
to be important, for instance, for  explaining a delayed cooling of the Earth's mantle \cite{Yuen:2001.1}.
There are also recent studies about non-Boussinesq contributions to convection
in binary-fluid mixtures where either a temperature dependent thermo-diffusion coefficient
or a dependence of the viscosity on the local composition was taken into account
\cite{Luecke:1987.1,Pleiner:2008.1,Glaessl:2010.1}.

Convection in binary-fluid mixtures is another classical, driven pattern-forming system
\cite{Legros:84,Turner:85.1}, which has attracted wide attention during the last decades
\cite{Legros:84,Turner:85.1,LandauVI,Brand:84.1,Cross:88.2,Knobloch:88,Zimmermann:89.5,Zimmermann:93.1,Luecke:1998.1}.
In  binary-fluid mixtures the concentration field of one of the two constituents enters
the basic equations as an additional dynamic quantity \cite{Legros:84,Turner:85.1,LandauVI}.  
Via the Soret effect ({\it thermophoresis}) a temperature gradient, that is applied vertically across
a convection cell, may cause variations of the concentration field which couples
into the Navier-Stokes equations for the velocity field via the buoyancy term.
Depending on the sign of the Soret effect, the heavier constituent is either driven
to the colder upper boundary or to the warmer lower boundary. In the former
case one obtains stationary convection patterns near the onset of convection and
in the latter case oscillatory patterns. Experimentally, the onset of convection
is well investigated in mixtures of alcohol and water as well as for $^3He/^4He$ mixtures \cite{CrossHo}.
The possibility of having both, a stationary as well as an oscillatory onset of convection,
including a so called codimension-2 bifurcation at the transition point, made it
to a very attractive model system for generic bifurcation phenomena \cite{Guckenheimer:83,CrossHo}. 

Colloidal suspensions, also known as nanofluids, may be considered as a further example
of a binary mixture, with the suspended particles being the second constituent.
Recently, convection in colloidal suspensions
\cite{Cerbino:2002.1,Cerbino:2005.1,Choi:2007,Cerbino:2009.1,Rehberg:2010} has been investigated experimentally
with a special focus on Soret driven convection \cite{Cerbino:2002.1,Cerbino:2005.1,Choi:2007},
on bistable heat transfer which is caused by sedimentation effects \cite{Cerbino:2009.1},
or on the effects of thermosensitive particles \cite{Rehberg:2010}.

Additionally, in several nanofluids with particle sizes in the range of $1-100$ nm a strong
dependence of the thermal conductivity on the concentration of nanoparticles was reported, see, e.g.,
Refs.~\cite{DasSK:2003.1,Eastman:2004.1,Prasher:2005.1,Buongiorno:2009.1,ChoiSUS:2009.1,Eapen:2010.1}.
This dependence presents another non-Boussinesq effect. In this work we investigate its impact on 
the onset of convection. Taking the particle-concentration-dependent thermal conductivity
into account, we go beyond other recent studies with the focus on convection in colloidal suspensions
\cite{Pleiner:2007.1,Nield:2010.2}, where the Boussinesq approximation has been used.

The enhancement of the thermal conductivity in fluids by increasing the concentration of
suspended nanoparticles was confirmed by a recent benchmark study with contributions of more than
thirty researchers \cite{Buongiorno:2009.1}. In spite of the promising applications of nanofluids
for improving heat transfer in cooling systems, a considerable number of open questions is left.
Among several possible transport phenomena, discussed to explain the heat transfer enhancement 
in nanofluids, Brownian diffusion and thermophoresis were identified as the two most important
ones \cite{Buongiorno:2006.1}. Both mechanisms build the crucial extensions from models
for a single component Newtonian fluid to models for convection in binary fluid mixtures
\cite{Legros:84,LandauVI}.

The paper is organized as follows: In Sec.~\ref{model} we briefly present the underlying
equations of motion and introduce a linear relation between the heat conductivity and 
the particle concentration, which leads to a nonlinear spatial dependence of the heat
conductive state. The essential methods of the linear stability analysis for the determination 
of the onset of convection are presented in Sec.~\ref{linstab}. Our numerical results
for the onset of convection include a prediction of an oscillatory onset even in the range
of positive values of the separation ratio for both cases, no-slip boundary conditions
in Sec.~\ref{no-slip} and free-slip boundary conditions in Sec.~\ref{free-slip}.
In Sec.~\ref{conclusions} we summarize and discuss our results.

\section{Basic equations and heat conductive state\label{model}}
To describe convection in a horizontal layer of a colloidal suspension
the common mean field approach for binary-fluid mixtures is used
\cite{Legros:84,Luecke:1998.1,Brand:84.1,Cross:88.2,Zimmermann:93.1,Guthkowicz:79}.
In addition, we take into account a linear dependence of the thermal
conductivity $\kappa$ on the mass fraction of the colloidal particles
$N({\bf r},t)$. The mass density of the colloidal particles $\rho_c$
is assumed to be similar to the mass density of the solvent $\rho_s$, i.~e. 
$\varepsilon = \rho_c / \rho_s \simeq 1\,.$
%
%
Furthermore, we assume small colloidal particles undergoing strong Brownian motion,
so that sedimentation effects become negligible. Deviations of  $N({\bf r},t)$
from the mean mass fraction $N_0$ may lead to a spatial dependence of the thermal
conductivity via the linear relation
\begin{align}
\kappa = \kappa_0 \left(1+ \gamma ~(N-N_0)\right)\,,\label{eq:Einstein} 
\end{align}
where $\kappa_0$ describes the mean thermal conductivity of the suspension and 
$\gamma = (1/\kappa_0)\,\partial\kappa / \partial N$ is a measure for the dependence of the thermal conductivity on the
concentration of the nanoparticles. Heat conduction experiments with small
volume fractions of nanoparticles match with the assumption $\gamma \sim 2.5$
\cite{DasSK:2003.1,Prasher:2005.1,Buongiorno:2009.1}.
 
The common set of basic transport equations for incompressible binary-fluid mixtures,
cf. Refs.~\cite{Legros:84,Brand:84.1,Cross:88.2,Zimmermann:93.1,Luecke:1998.1}, involves
the temperature field $T(\vec{r},t)$, 
the mass fraction of the particles $N(\vec{r},t)$,
the fluid velocity $\vec{v}(\vec{r},t)$, 
the density of the mixture $\rho(\vec{r},t)$, 
and the pressure field $p(\vec{r},t)$:
\begin{subequations}
\begin{align}
\nabla \cdot \vec{v} &= 0 \,, \label{eq:incompressible} \\
(\partial_t +  \vec{v} \cdot \nabla) \,T &= \nabla \cdot \left( \chi \nabla T \right)\,,  \label{eq:heat} \\ 
(\partial_t +  \vec{v} \cdot \nabla) \,N &= D \,\nabla \cdot  \left(\nabla  N + \frac{k_T}{T} \,\nabla  T \right)\,, \label{eq:continuity} \\
(\partial_t +  \vec{v} \cdot \nabla) \,\vec{v} &= -\frac{1}{\rho_0}\,\nabla  p + \nu \, \Delta  \vec{v} 
+  \frac{ \rho}{\rho_0} \,\vec{g}\,. \label{eq:navier-stokes}
\end{align}
\label{eq:basic}
\end{subequations}
Eq.~\eqref{eq:incompressible} describes the incompressibility of the fluid.
$\chi = \kappa / \rho_0$ in the heat equation \eqref{eq:heat} denotes
the thermal diffusivity of the mixture, which is in our model a function
of the concentration of the suspended colloids. $D$ in Eq.~\eqref{eq:continuity}
is the diffusion constant, which takes due to the size of the colloidal particles
much smaller values than in molecular binary-fluid mixtures. The dimensionless
thermal-diffusion ratio $k_T$ representing the cross coupling between the temperature
gradient and the particle flux is related to the Soret coefficient $S_T$ via 
$k_T / T =  N(1- N) S_T$, which can be either positive  or negative. Throughout
this work $k_T / T \simeq N_0(1- N_0) S_T$ is regarded as constant. $\nu$  in
the Navier-Stokes-equations \eqref{eq:navier-stokes} is the kinematic viscosity.
The gravity field $\vec{g} = - g \vec{e}_z$ is chosen parallel to the z-direction.

For the local density $\rho$ of the suspension we use a linearized equation of state 
\cite{Gershuni:76,Legros:84},
\begin{align}
 \rho = \rho_0 \left[1-\alpha(T-T_0) + \beta(N-N_0) \right]\, ,
\end{align}
with the thermal expansion  coefficient $\alpha = - (  1/\rho_0) \partial \rho /\partial T$
and $ \beta = (1/\rho_0) \partial \rho/\partial N$ reflecting the density contrast between
the solvent and the suspended particles. According to the Boussinesq approximation
this dependence of the density is taken into account only within the buoyancy term.
The sign of $\beta$ indicates whether the colloidal particles have a higher or a lower mass density 
compared to the solvent. Here, we assume $\beta >0$ corresponding to $\varepsilon \gtrsim 1$.

{\it Boundary conditions.~}
The  fluid layer is confined between two impermeable, parallel plates
at a distance $d$, and extends infinitely in the ($x$-$y$)-plane. 
The lower plate at $z=-d/2$ is kept at a higher temperature $T_0 + \delta T/2$
than the upper plate at $z=+d/2$ with  the lower temperature $T_0 - \delta T/2$.
Together with a vanishing mass current at the boundaries and realistic no-slip
conditions for the flow field we have the following set of boundary conditions
at $z=\pm d/2$:
\begin{subequations}
\begin{align}
 T&= T_0 \mp \frac{\delta T}{2}\,, \label{eq:boundary3} \\
0 &= \partial_z  N + \frac{k_T}{T_0} ~ \partial_z  T,  \label{eq:boundary2}\\
0 &=  v_x =  v_y=  v_z= \partial_z  v_z  \,. \label{eq:boundary1}
\end{align}
\end{subequations}
For geophysical applications free-slip boundary conditions 
\begin{align}
 0 &=  \partial_z v_x =  \partial_z v_y=  v_z= \partial_z^2  v_z  \,,
\end{align}
for the flow field are often considered to be more realistic \cite{OgawaM:1991.1}. 
In the case of a constant thermal conductivity and free-slip, permeable boundary conditions
an analytical determination of the onset of convection is possible \cite{Brand:84.1}.
By introducing a concentration dependent thermal conductivity this advantage is lost 
and one has to rely on numerical methods.

\subsection{Heat conductive state} 
%
Together with the concentration dependent heat conductivity given by
Eq.~(\ref{eq:Einstein}) the constant vertical heat current
$j_z = -\chi \partial_z T_{\, \rm{cond}}$ leads to a nonlinear
$z$-dependence of the time-independent temperature distribution
$T_{\, \rm{cond}}(z)$ and the corresponding concentration distribution
$N_{\, \rm{cond}}(z)$ of the heat conductive state. Both are derived in
the Appendix \ref{app1} and are of the form
\begin{subequations}
\label{heatcond}
\begin{align}
\label{eq:condT}
 T_{\, \rm{cond}} &= T_0 + \frac{\delta T}{\xi} \left( \frac{1}{2} \left( 1+Y(\xi) \right) - W(z,\xi) \right)\\
\label{eq:condT_linear}
 &=T_0 - \delta T\,\frac{z}{d}-\xi\,\frac{\delta T}{2}\left(\frac{1}{4}-\frac{z^2}{d^2} \right) + \mathcal{O}(\xi^2)\,,\,\,\,
\\
\label{eq:condN}
 N_{\, \rm{cond}} &= N_0 + \frac{\delta\! N}{\xi} \Big( 1- W(z,\xi) \Big) \\
\label{eq:condN_linear}
&= N_0 - \delta\! N \frac{z}{d} -\xi\,\frac{\delta\! N}{2}\left(\frac{1}{12}-\frac{z^2}{d^2}\right)+ \mathcal{O}(\xi^2)\,,
\end{align}
\label{eq:groundstate}
\end{subequations}
with the abbreviations
\begin{subequations}
\begin{align}
\label{W(z)}
 W(z,\xi) &= \sqrt{\xi \left( 1+Y(\xi) \right) \frac{z}{d} + \frac{1}{2} \left( 1+ Y(\xi) + \frac{1}{3} \xi^2 \right) } \,,\\
\label{Y(xi)}
 Y(\xi) &= \sqrt{1-\frac{1}{3} \xi^2} \, , \, \, \, \,  \qquad 
 \xi = \gamma \, k_T \, \frac{\delta T}{T_0} \,
\intertext{and}
\delta \!N &= -\frac{k_T}{T_0}\,\delta T\,.
\end{align}
\end{subequations}
For finite values of $\gamma$ the vertical heat current,
\begin{align}
j_z = \frac{\chi_0 \delta T}{2d} \left( 1+\sqrt{1-\frac{1}{3}\xi^2} \right)\, ,
 \label{heatcurrent}
\end{align}
is reduced compared to the situation of a constant heat conductivity
being independent of the particle concentration.

We would like to stress that Eqs.~(\ref{heatcond}) are derived under
the assumption of small values of $\delta N$, but these formulas are
still reasonable for values up to $\delta N < N_0/3$. The restriction
$|\xi|< \sqrt{3}$ according to Eq.~(\ref{Y(xi)}) has the same origin and
is fulfilled by all parameters chosen in this work. For stronger variations
of $\delta N$ the assumption of a constant thermodiffusion coefficient 
$k_T/T \simeq N_0(1-N_0)S_T$ is not justified anymore and a generalized 
approach has to be chosen.

\subsection{Dimensionless equations of convective fluid motion}
%
For the further analysis it is convenient to separate the basic heat conductive state in
Eq.~(\ref{heatcond}) from convective contributions as follows:
$ T(\vec{r},t) =  T_{\, \rm{cond}}(z) +  T_1(\vec{r},t)$ and $ N(\vec{r},t) =  N_{\, \rm{cond}}(z) +  N_1(\vec{r},t)$. 
Using the rotational symmetry in the fluid layer 
we can restrict our analysis to two spatial dimensions, namely to the ($x$-$z$)-plane.
With this simplification the fluid velocity $\vec{v}=(v_x,0,v_z)$ can be expressed
by a {\it stream function} $\phi(x,z,t)$:
\begin{align}
 v_z = \partial_x \phi ~,\quad v_x = -\partial_z \phi\,.
\end{align}
Subsequently all lengths are scaled by the vertical distance $d$ and
times by the vertical thermal diffusion time $d^2/\chi_0$.
Scaling the temperature field $T$ by $(\chi_0 \nu_0) / (\alpha g d^3)$,
the concentration field $N$ by $-(k_T \chi_0 \nu_0) / (T_0 \alpha g d^3)$
and the stream function $\phi$ by $\chi_0 d$
we are left with five dimensionless parameters:
The {\it Rayleigh number} $R$, the {\it Prandtl number} $P$, the {\it Lewis number} $L$, and the
{\it separation ratio $\psi$},
\begin{align}
\label{scaledef}
 P    = \frac{\nu_0}{\chi_0},\quad  
 L    = \frac{D}{\chi_0},\quad 
 R    = \frac{\alpha g d^3}{\chi_0 \nu_0}\, \delta  T,\quad  
 \psi = \frac{\beta k_T}{\alpha T_0}\,,
\end{align}
are well known from molecular binary-fluid mixtures \cite{Cross:88.2,Zimmermann:93.1}.
The fifth dimensionless quantity,
\begin{align}
 \zeta = \gamma \,\frac{\nu_0 \chi_0}{g \beta d^3}\,,
\end{align}
is introduced to characterize the spatially varying contribution to the thermal diffusivity
caused by the concentration-dependence of the thermal conductivity of the suspension. 
An illustration of its physical meaning is obtained by considering the thermal conductivity
contrast between the upper and the  lower boundary,
\begin{align}
 {\tilde \kappa} &= \frac{1 + \frac{1}{2} \, \gamma \, k_T \, \delta T / T_0}
                       {1 - \frac{1}{2} \, \gamma \, k_T \, \delta T / T_0} =
\frac{1+\frac{1}{2} \, R \, \psi \, \zeta}{1-\frac{1}{2} \, R \, \psi \, \zeta} \, ,
 \label{kontrast}
\end{align}
which is essentially a function of the product of
the three dimensionless control parameters  $R \, \psi \, \zeta = \xi$.

In the following we will discuss our results essentially in dependence on
$\psi$ and $\zeta$, whereas $P$ and $L$ are considered as constants.

Introducing a rescaled temperature deviation $\theta = (R / \delta T) T_1$,
a rescaled concentration deviation $\tilde N_1 =-(T_0 R / k_T \delta T)N_1$,
and a rescaled stream function $\Phi = 1/(\chi_0 d)\,\phi$
in  terms of these dimensionless quantities and using the combined function
$\tilde c = \tilde N_1 - \theta$ instead of $\tilde N_1$ we obtain:

\begin{subequations}
\label{scaleeq}
\begin{align}
 \partial_t \theta &- W_0 \Delta \theta - R \, \psi \, \zeta W_1(\partial_z c +2\partial_z \theta) \\ \nonumber
                   &- W_2(c+\theta) - R W_1 \partial_x \Phi = \\ \nonumber
                   &- \psi \zeta \left[(\partial_x \theta)\partial_x (\theta+c) +(\partial_z \theta)\partial_z (\theta+c) + \Delta \theta(\theta+c)\right] \\ \nonumber
                   &+ (\partial_z \Phi \partial_x - \partial_x \Phi \partial_z) \, \theta \, ,\\[1mm] 
 \partial_t c      &+ W_0 \Delta \theta + R \, \psi \, \zeta W_1(\partial_z c +2\partial_z \theta) \\ \nonumber
                   &+ W_2(c+\theta) - L \Delta c  =  \\ \nonumber
                   &+ \psi \zeta \left[(\partial_x \theta)\partial_x (\theta+c) +(\partial_z \theta)\partial_z (\theta+c) + \Delta \theta(\theta+c)\right] \\ \nonumber 
                   &+ (\partial_z \Phi \partial_x - \partial_x \Phi \partial_z) \, c \, ,\\[1mm]
\partial_t \Delta \Phi &- P \Delta^2 \Phi - P (1+\psi) \partial_x \theta - P \psi \partial_x c = \\ \nonumber
                   & \quad \, P (\partial_z \Phi \partial_x - \partial_x \Phi \partial_z) \, \Delta \Phi  \, ,
\end{align}
\end{subequations}
with the abbreviations
\begin{subequations}
\begin{align}
 W_0 &=                          W(z,\xi) \, ,\\
 W_1 &= \frac {1}{\xi}\partial_z W(z,\xi) \, ,\\
 W_2 &=             \partial_z^2 W(z,\xi) \, .
\end{align}
\label{Ws}
\end{subequations}

For reasons of simplicity all the tildes have been suppressed.
No-slip, impermeable boundary conditions for the fields $\theta$, $c$, and $\Phi$ demand
\begin{align}
\label{bcsc}
 \theta = \partial_z c =\Phi=\partial_z \Phi=0 \quad \text{at $z= \pm\frac{1}{2}$}
\end{align}
while free-slip, permeable boundary conditions \cite{Brand:84.1,Guthkowicz:79}
\begin{align}
\label{bcsc2}
 \theta = c =\Phi=\partial_z^2 \Phi=0 \quad \text{at $z= \pm\frac{1}{2}$}\, .
\end{align}
In this work we present both, results based upon no-slip, impermeable boundary conditions and 
results based upon free-slip, permeable boundary conditions. 

\section{Linear stability of the heat conductive state and onset of convection\label{linstab}}
In order to determine the parameters at the onset of convection we
investigate the linear stability of the heat conductive state given
by Eqs.~(\ref{heatcond}) with respect to small perturbation fields:
$\,\theta (x,z,t)$, $c(x,z,t)$, and $\Phi(x,z,t)$. The reduced set
of three linear PDEs with constant coefficients may be solved by the ansatz
\begin{align}
\label{ansatz1}
\left(\begin{array}{c}
  \theta(x,z,t)\\
  c(x,z,t)\\ 
  \Phi(x,z,t) 
\end{array}
\right) 
&=\vec{u}_0(z) ~e^{iqx} e^{\sigma t}+ c.c.\,,
\end{align}
with the vector function
\begin{align}  \vec{u}_0(z) &=
\left(
\begin{array}{c} 
  \bar\theta(z) \\ 
  \bar c(z) \\
  \bar \Phi(z)/(iq) 
\end{array} 
\right)\,
\end{align}
and $c.c.$ denoting the complex conjugate, leading to a boundary eigenvalue problem
with respect to $z$ and the eigenvalue $\sigma$. We solve the remaining
linear ODEs by two different methods as summarized in the following paragraphs.

The first one is the standard {\it shooting method} as described in detail
for binary-fluid convection in Ref.~\cite{Zimmermann:93.1}. The resulting
coupled ODEs for the components of the vector function ${\bf u}_0(z)$
are integrated for a set of initial conditions at one boundary. With
the value of ${\bf u}_0$ at the opposite boundary a determinant
$f(\sigma,R,q,Q,P,L,\psi)$ follows. Keeping the initial conditions fixed, 
either $R$ or $\sigma$ are varied such that $f$ vanishes. The resulting
values of $\sigma$ and $R$ are functions of the remaining parameters.

The second approach is based upon the so-called {\it Galerkin method}. The components
of ${\bf u}_0(z)$ are expanded with respect to a suitable chosen set of functions
fulfilling already the boundary conditions, i.~e. either Eqs.~\eqref{bcsc} or
Eqs.~\eqref{bcsc2}. Examples for this alternative numerical method may be found
in Refs.~\cite{Pesch:1996.1,Busse:1974.2,Canuto:1987.1}. The resulting
generalized algebraic eigenvalue problem is solved numerically.

At the onset of convection the small perturbations $\theta$, $c$, and $\Phi$
neither grow nor decay. This is the so called neutral stability condition,
where the real part $\mbox{Re}(\sigma)$ of the  eigenvalue $\sigma$ with
the largest real part (fastest growing mode) vanishes:
\begin{align}
\label{neutcond}
\mbox{Re}(\sigma)=0\, \quad \text{with}\quad \sigma=\sigma(R,q,\zeta,P,L,\psi)\,.
\end{align}
This yields the Rayleigh number 
\begin{align}
\label{neutcurve}
 R_0(q)=R_0(q,\zeta,P,L,\psi)
\end{align}
as a function of the chosen wave number $q$ and describes the so-called
{\it neutral curve} with a minimum at the critical wave number $q_c$ and the
critical Rayleigh number $R_c=R_0(q_c)$. Convection sets in with a wave number
$q \simeq q_c$ by crossing $R_c$ from below.
Depending on parameters, the onset of convection may take place via a 
{\it stationary bifurcation} with a vanishing imaginary part of the eigenvalue,
$\mbox{Im}(\sigma) = 0$, or via a {\it Hopf bifurcation} with a finite
Hopf frequency $\mbox{Im}(\sigma) = \pm \, \omega_0(q,\zeta,P,L,\psi)$ with
its critical value $\omega_c=\omega_0(q_c)$.

Throughout this work we choose for reasons of simplicity the Prandtl number  $P=10$. 
Since nanoparticles are much larger than, for instance, alcohol molecules in water,
their mass diffusion is more than two orders of magnitude smaller. Accordingly,
the Lewis number in nanofluids takes considerably smaller values of about $L=10^{-4}$
\cite{Piazza:2008.2}. Growing linearly with the particle's size \cite{Piazza:2008.2}
the Soret effect can be changed in a wide range by varying the mass density with 
respect to the base fluid or the particle diameter. For small nanoparticles in water
$\zeta \simeq 0.01$ and in Glycerin $\zeta \simeq 10$ are appropriate values.

\subsection{No-slip, impermeable boundary conditions\label{no-slip}} 
%
It is a major result of this work that a particle-concentration-dependent thermal conductivity,
described by finite values of $\zeta$, leads in the range  $\psi >0$ to a shift
of the onset of convection to larger values of $R_c$. As another important result
we find in the range $\psi \simeq L$ and beyond some critical value $\zeta_c$
an exchange of instabilities from a stationary bifurcation to an oscillatory one
and characterize this transition as a function of $\psi$ and $\zeta$.

\begin{figure}[ttt]
  \subfigure[\, \mbox{Neutral curves for different $\zeta$ at}~$\, \psi=10^{-5}$.]
  {
  \includegraphics[width=0.9\columnwidth]{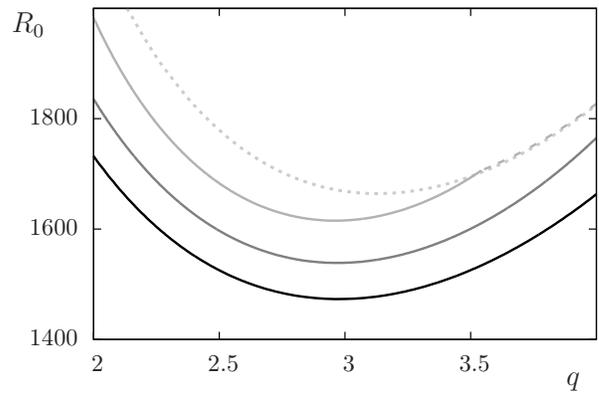}
  }
  \vspace{2mm}
  \subfigure[\, \mbox{Neutral curves near codimension-2 bifurcation.}]
  {
  \includegraphics[width=0.9\columnwidth]{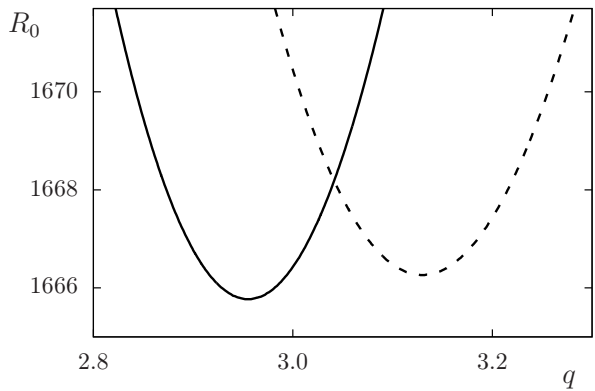}
  }
\vspace{-2mm}
  \caption{\label{neutstationary1}
           Neutral curves $R_0(q)$ are shown for  different values 
           of the conductivity parameter $\zeta$ and $\psi=10^{-5}$.
           Solid lines mark stationary bifurcations and dotted/dashed lines oscillatory ones.
           In part (a) one has $\zeta=0$ (black solid line), $\zeta=0.06$ (dark-gray solid line), 
           $\zeta = 0.08$ (gray solid and dashed lines),  and  $\zeta = 0.10$ (light-gray dotted line).
           In part (b) two neutral curves, that correspond to the two eigenvalues with the largest real parts,
           are shown at $\zeta = 0.0875$. Again, the solid line
           indicates a stationary bifurcation and the dashed line an oscillatory one.
}

\end{figure}

\begin{figure}[ttt]
  \subfigure[\, Neutral curves at $\, \psi=10^{-4}$.]
  {
  \includegraphics[width=0.9\columnwidth]{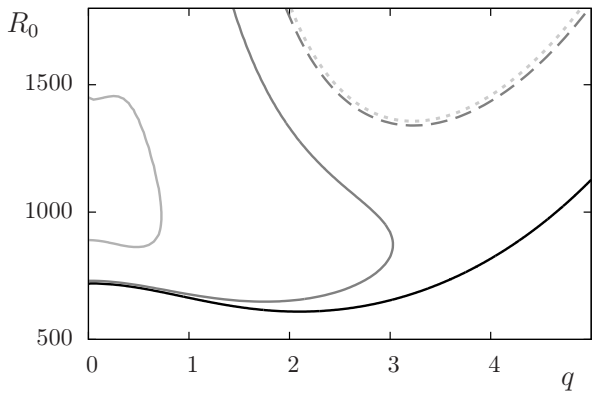}
  }
\vspace{-2mm}
  \caption{\label{neutstatzeta_2}
           Neutral curves $R_0(q)$ are shown for $\psi=10^{-4}$ and the heat
           conductivity parameter $\zeta=0$ (black solid line),
           $\zeta=0.05$ (dark-gray lines), $\zeta=0.15$ (gray solid line),
           and $\zeta=0.20$ (light-gray dotted line). At $\zeta=0.05$ the neutral curve splits
           into the dark-gray solid part for the stationary and the dark-gray dashed line
           for the oscillatory instability. At $\zeta=0.15$ the gray solid line
           marks the stationary branch and the oscillatory one is not shown.
           The light-gray dotted line corresponds to the oscillatory instability at $\zeta=0.20$.
          }
\end{figure}

The neutral curve $R_0(q)$ belonging to the lowest bifurcation from
the heat conductive state at different values of $\zeta$ is shown
in Fig.~\ref{neutstationary1} and Fig.~\ref{neutstatzeta_2} for two
representative positive values of the separation ratio $\psi$,
namely for $\psi=10^{-5} \ll L$ (Fig.~\ref{neutstationary1}) and
$\psi=10^{-4} = L$ (Fig.~\ref{neutstatzeta_2}). In both Figures
the black solid line describes the neutral curve $R_0(q)$ corresponding
to molecular binary fluids, i.~e., $\zeta = 0$. It is included for 
illustrating the relative changes of the neutral curve as a function of $\zeta$. 
We also mention for completeness that the Rayleigh number at the onset of
convection in binary fluids decreases  in the range of $\psi >0$ with
increasing values of $\psi$ starting at $R_c (\psi=0) \simeq 1708$.
Simultaneously the critical wave number $q_c$ of the stationary bifurcation 
tends to zero in the range $\psi \gtrsim L$ \cite{Cross:88.2,Zimmermann:93.1,Luecke:1998.1}.

For increasing values of  $\zeta$ the neutral curve $R_0(q)$ of the stationary
bifurcation is shifted to larger values, as illustrated in Fig.~\ref{neutstationary1}(a).
For $\zeta=0.08$ at about  $q \simeq 3.55$ along $R_0(q)$ a transition takes place
from a stationary bifurcation in the range $q \le 3.55$ to an oscillatory one
in the range $q \gtrsim 3.55$. We mention that here and in all following Figures
solid lines mark stationary bifurcations whereas dashed, dotted, or dashed-dotted curves 
indicate oscillatory bifurcations.
The neutral curve is determined by the condition given in Eq. (\ref{neutcond}) for the
eigenvalue with the largest real part. At about $q \simeq 3.55$ the real parts of three
different eigenvalues cross each other as a function of $q$, where the eigenvalue of the
stationary bifurcation has the largest eigenvalue in the range $q \le 3.55$ and a pair of
complex conjugate eigenvalues leading to an oscillatory bifurcation has the largest real part
in the range $q \gtrsim 3.55$. Accordingly, at the transition between the two instabilities 
one has an unsteady change of the slope along the lowest neutral line. At a slightly larger
value of $\zeta$ this transition takes place at a smaller value of $q$ and the path along
the lowest parts of the two neutral curves belonging to the two instabilities becomes
nonmonotonic as illustrated in Fig.~\ref{neutstationary1}(b) for $\zeta=0.0875$.
At the crossing point of the two neutral curves in  Fig.~\ref{neutstationary1}(b)
the eigenvalue of the stationary branch vanishes as well as the real parts
of the two complex conjugate eigenvalues at the oscillatory branch.
This type of an exchange of instabilities is different from the well known
codimension-2 point in molecular binary fluids \cite{Cross:88.2,Zimmermann:93.1}, 
where only two of the three eigenvalues are involved and one has a so-called double 
zero eigenvalue problem \cite{Guckenheimer:83}.

At larger, but still small values of the separation ratio $\psi$
the transition from a stationary to a Hopf bifurcation, which occurs by
increasing $\zeta$, takes place in a different and
more diverse manner, as exemplified in Fig.~\ref{neutstatzeta_2}
for $\psi=10^{-4}$. The neutral curve of the stationary bifurcation
at $\zeta=0$ (black solid line) is strongly deformed by changing $\zeta$
to $\zeta=0.05$ (dark-gray curves). The minimum of the latter curve
at $q_c$ is shifted to a smaller value compared to the black line.
Moreover, for certain wave numbers, e.g. $q=2.5$, one eigenvalue $\sigma$
vanishes at two different values of $R$ by crossing two times the
dark-gray solid line. Between these two values of $R$ the growth rate
$Re(\sigma)$ is positive, otherwise negative. Beyond the dark-gray solid curve 
at an even higher value for $R$ at $q=2.5$ and $\zeta=0.05$ the real parts
of a pair of complex conjugate eigenvalues change their sign and become positive,
namely beyond the dark-gray dashed line. Increasing the conductivity parameter
up to $\zeta=0.15$, the gray solid line for the stationary bifurcation
results, which is even further deformed compared to the case $\zeta = 0.05$. 
Increasing $\zeta$ beyond $\zeta=0.15$ the stationary branch vanishes
at about $\zeta=0.165$ and beyond this value only a Hopf bifurcation
from the heat conductive state takes place as shown by the light-gray dotted line
for $\zeta=0.2$. 
Since the value of the Rayleigh number of the neutral curves at the Hopf branch changes 
only slightly as a function of $\zeta$, the Hopf branch belonging to $\zeta = 0.15$ is not 
shown.

\begin{figure}[ttt]
  \subfigure[\, Critical Rayleigh number $R_c(\zeta)$.]
  {
  \includegraphics[width=0.9\columnwidth]{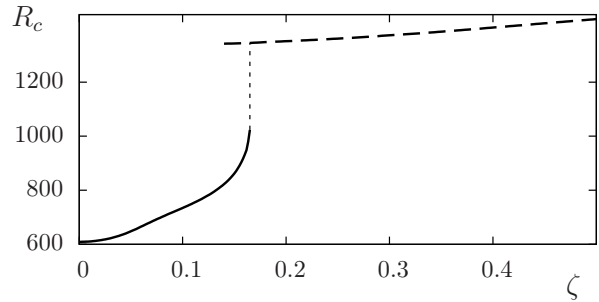}
  }
 \vspace{2mm}
  \subfigure[\, Critical wave number $q_c(\zeta)$.]
  {
  \includegraphics[width=0.9\columnwidth]{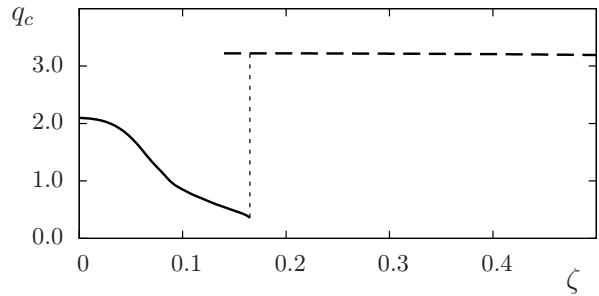}
  }
 \vspace{2mm}
  \subfigure[\, Critical frequency $\omega_c(\zeta)$.]
  {
  \includegraphics[width=0.9\columnwidth]{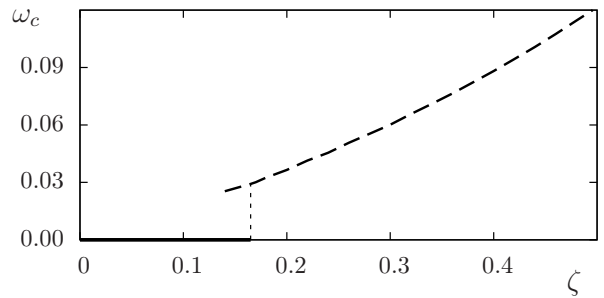}
  }
\vspace{-5mm}
\caption{\label{critvalues}
         Part (a) shows the critical Rayleigh number $R_c$,
         (b) the critical wave number $q_c$, and (c) the critical frequency
         $\omega_c$ as a function of $\zeta$ for $\psi = 10^{-4}$.
         Solid lines mark a stationary onset of convection and
         dashed lines an oscillatory one.
}
\end{figure}

This transition scenario, that increasing values of $\zeta$  lead to a growth
of the  critical Rayleigh number at the onset of stationary convection
and finally to a Hopf bifurcation to convection at even higher Rayleigh numbers, 
is illustrated from a slightly different perspective by Fig.~\ref{critvalues}.
In this Figure the critical values $R_c=R_0(q_c)$, $q_c$ and $\omega_c$
at the minimum of the lowest neutral curves are plotted as functions
of the conductivity parameter $\zeta$. The solid lines, describing the 
stationary bifurcation as a function of $\zeta$, cease to exist at about $\zeta=0.165$.
During the process of disappearance of the stationary bifurcation
the neutral curve of this bifurcation is deformed as illustrated
in Fig.~\ref{neutstatzeta_2}. Beyond  $\zeta=0.165$ a Hopf bifurcation
from the heat conductive state is preferred.
Within the oscillatory region, $R_c$ varies only slightly with $\zeta$
and takes values, that are more than $100 \%$ larger than the corresponding
value for $\zeta = 0$. The critical wave number decreases initially,
exhibits a major jump at the transition point and remains finally 
nearly constant, taking a value of $q_c \simeq 3.2$, which is close
to the critical wave number for a simple Newtonian fluid, $q_c^{N\!F} = 3.116$.
The critical Hopf frequency is monotonically increasing.

\begin{figure}[ttt]
  \subfigure[\, Critical Rayleigh number $\, R_c(\psi)$.]
  {
  \includegraphics[width=0.9\columnwidth]{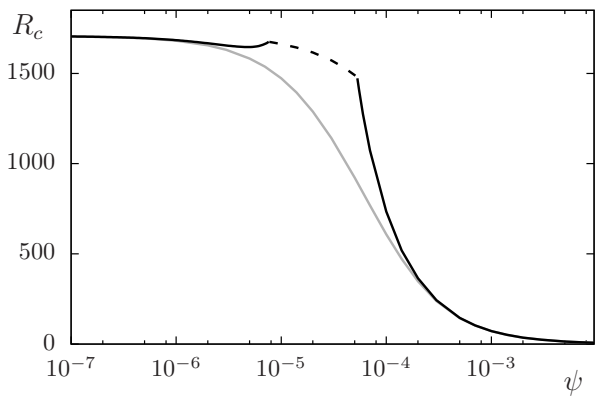}
  }
\vspace{-2mm}
  \caption{\label{RvsPsi_1}
           The critical Rayleigh number $R_c$ is given as a function
           of $\psi$ for $\zeta = 0.0$ (gray line) and
           $\zeta = 0.1$ (black lines). In the latter case the solid
           line marks the stationary and the dashed line the Hopf branch.
           At the transition from the stationary to the Hopf branch one has 
           a codimension-2 point.
}
\end{figure}

\begin{figure}[ttt]
  \subfigure[\, Critical Rayleigh number $\, R_c(\psi)$.]
  {
  \includegraphics[width=0.9\columnwidth]{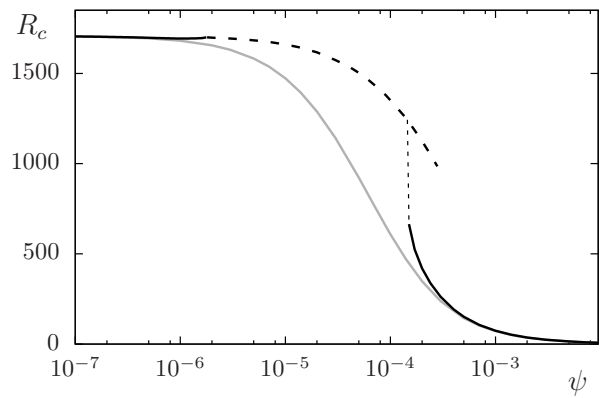}
  }
  \vspace{2mm}
  \subfigure[\, Critical wave number $\, q_c(\psi)$.]
  {
  \includegraphics[width=0.9\columnwidth]{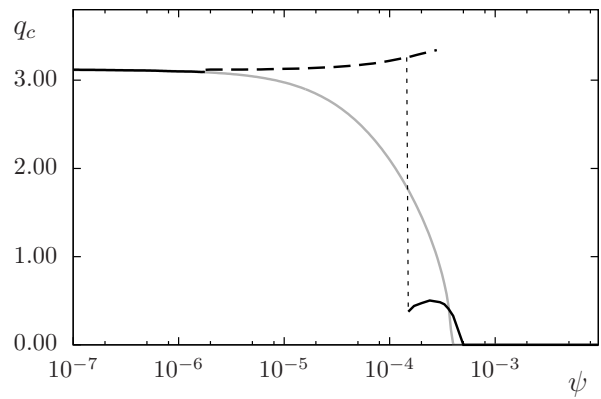}
  }
\subfigure[\, Critical Hopf frequency $\, \omega_c(\psi)$.]
  {
  \includegraphics[width=0.9\columnwidth]{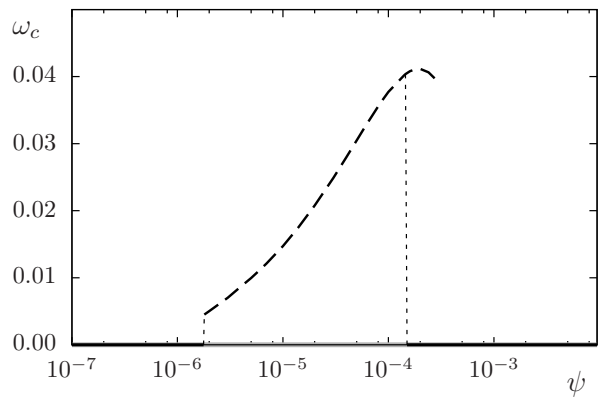}
  }
\vspace{-2mm}
  \caption{\label{RvsPsi}
           In part (a) the critical Rayleigh number $R_c$ is shown as a function
           of $\psi$ for $\zeta = 0$ (gray line) and $\zeta = 0.2$ (black lines),
           whereby in the latter case the solid lines mark a stationary bifurcation
           and the dashed line a Hopf bifurcation. Part (b) shows the corresponding
           critical wave number $q_c (\psi)$ and part (c) the frequency $\omega_c (\psi)$.
}

\end{figure}

A third perspective on the transition from a stationary bifurcation to
a Hopf bifurcation is provided by Fig.~\ref{RvsPsi_1} for $\zeta=0.1$
and Fig.~\ref{RvsPsi} for $\zeta=0.2$. In both Figures we present
the critical values $R_c$ resp. $R_c$, $q_c$ and $\omega_c$ as functions
of the separation ratio $\psi$. As a guide to the eye we have also 
included the well known limiting case for $\zeta=0$ (gray solid lines).
Again, solid curves describe the critical values for a stationary
bifurcation, if this threshold is lower than the Hopf bifurcation,
and dashed lines the critical values of the Hopf bifurcation.

The left transition from a stationary bifurcation to a Hopf bifurcation 
in Fig.~\ref{RvsPsi_1} and Fig.~\ref{RvsPsi} is a codimension-2 bifurcation
as illustrated in  Fig.~\ref{neutstationary1}(b). The critical Rayleigh numbers at
the minimum of the neutral curve of the stationary bifurcation at $q_c^S$ and at
the minimum of the Hopf branch at $q_c^H$ come rather close to each other with
$q_c^H-q_c^S \simeq 0.025$ in Fig.~\ref{RvsPsi}.

With further increasing values of $\psi$ the transition from the Hopf branch 
back to a stationary branch is different for $\zeta=0.1$ (Fig.~\ref{RvsPsi_1})
and $\zeta=0.2$ (Fig.~\ref{RvsPsi}). The right transition in Fig.~\ref{RvsPsi_1}
is similar to the left one and resembles qualitatively the scenario shown
in Fig.~\ref{neutstationary1}(b). Reducing $\zeta$ below $\zeta=0.1$ 
the two transitions shown in Fig.~\ref{RvsPsi_1} approach each other and one
obtains a rather degenerate bifurcation structure. The interesting nonlinear
dynamics in this parameter range will be discussed in more detail elsewhere.
The right transition in Fig.~\ref{RvsPsi} shows a jump in the wave number and
is in this respect rather different: By approaching this transition from larger
values of $\psi$ the stationary branch ceases to exist (c.f. Fig.~\ref{neutstatzeta_2}).
To the right of this second transition point convection sets in stationary. However, 
with decreasing values of $\psi$ the stationary part of the neutral curve becomes more 
and more deformed (just like in the case of increasing values of $\zeta$ at a fixed $\psi$, 
cf. Fig.~\ref{neutstatzeta_2}) and finally ceases to exist at the codimension-2 point.
Also this transition with quite disparate wave numbers at the two bifurcations
bears an interesting nonlinear behavior to be discussed elsewhere.

\begin{figure}[ttt]
  \includegraphics[width=0.9\columnwidth]{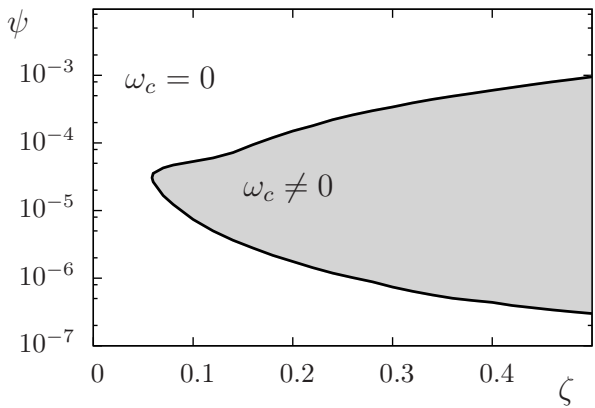}
\vspace{0mm}
  \caption{\label{Psivszeta}
          Location of the two codimension-2-points (cf. Fig.~\ref{RvsPsi_1} and
          Fig.~\ref{RvsPsi}) in the ($\zeta$-$\psi$)-plane. In the gray region inside
          the black curve the onset of convection takes place via a Hopf bifurcation
          and otherwise via a stationary bifurcation.
}
\end{figure}

The $\psi$-range where convection sets in via a Hopf bifurcation changes
as a function of the conductivity parameter $\zeta$ as shown in Fig.~\ref{Psivszeta}.
In the gray region within the displayed curve we find a Hopf bifurcation to convection,
outside a stationary bifurcation takes place. Close to the left nose of this curve
one has two transition points as in Fig.~\ref{RvsPsi_1}, whereas for larger values
of $\zeta$ the situation resembles the one shown in Fig.~\ref{RvsPsi},
with a strong jump in the wave number and the Rayleigh number along
the upper part of the curve.

In the limit of a vanishing Soret effect, i.~e., $\psi \rightarrow 0$, temperature gradients
do not cause concentration gradients anymore, the particle concentration becomes
homogeneous and the heat diffusivity $\chi$ is independent of the temperature.
In molecular binary-fluid mixtures one has a codimension-2 point at $\psi_{\rm{CTP}} \simeq -L^2$, 
with a stationary bifurcation for $\psi > \psi_{\rm{CTP}}$ and a Hopf bifurcation in the range
$-1 < \psi < \psi_{\rm{CTP}}$ \cite{Brand:84.1,Cross:88.2,Zimmermann:93.1}. In colloidal
suspensions with $L \simeq 10^{-4}$ or smaller this codimension-2 point $\psi_{\rm{CTP}}$
is even closer to zero.

In the range $\psi <  \psi_{\rm{CTP}}$ the Soret effect causes an enhancement
of the particle concentration near the lower plate and with $\gamma>0$ also the
heat conductivity increases at this boundary. For rising values of $\zeta$ we find 
a monotonous reduction of the critical Rayleigh number at the Hopf bifurcation
compared to the limit $\zeta = 0$. This is an opposite trend in comparison to
the enhancement of the threshold found in the range $\psi > 0$. 
However, this reduction of the Rayleigh number at the Hopf bifurcation is rather small and no 
deformation of neutral curves takes place. 
With increasing $\zeta$ the critical wave number $q_c$ decreases slightly and the frequency $\omega_c$
increases almost linearly, similar to the behavior depicted in Fig.~\ref{critvalues} (c).
For example, at $\psi = {-10}^{-4}$ and in the range $0\leq\zeta\leq0.5$ $R_c$ decreases with rising values 
of $\zeta$ from $R_c = 1708$ at $\zeta = 0$ to $R_c = 1624$ at $\zeta = 0.5$, $q_c$ decreases from 
$q_c = 3.116$ to $q_c = 3.097 $ and $\omega_c$ increases from  $\omega_c = 0.19$ to $\omega_c = 0.28$.
 
The thermal conductivity contrast $\tilde \kappa$ is a function of the product $R \psi \zeta = \xi$.
Since $R_c$ keeps large values at the Hopf bifurcation with decreasing $\psi$, in contrast
to the range $\psi>0$ \cite{Cross:88.2,Zimmermann:93.1}, the assumption of a linear
variation of the particle concentration along the vertical $z$ coordinate is violated
at medium values of $\zeta$, i.~e., the linear dependence of the thermal conductivity
on the particle concentration becomes violated for smaller values of $\zeta$ than
in the range $\psi>0$.

\begin{figure}[ttt]
  \subfigure[\, Stationary convection
]
  {
  \includegraphics[width=0.9\columnwidth]{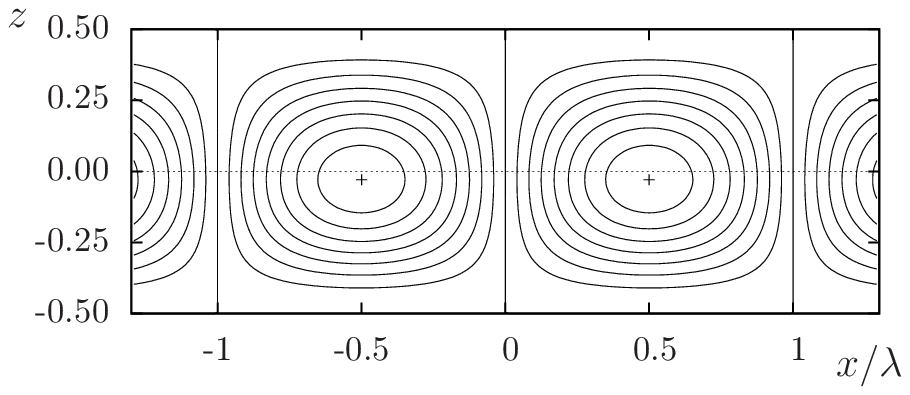}
  }
  \vspace{2mm}
  \subfigure[\, Left travelling wave ($\omega_c > 0$)
]
  {
  \includegraphics[width=0.9\columnwidth]{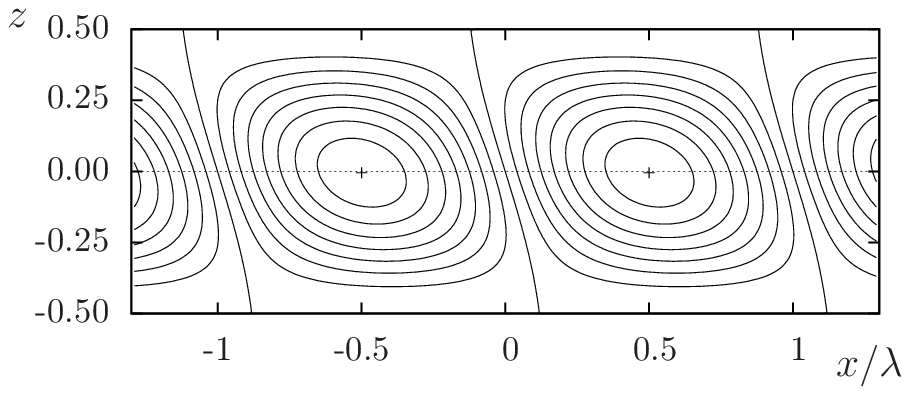}
  }
\vspace{-2mm}
  \caption{\label{flowfield}
        Contour lines of the stream function $\Phi$ at the onset of convection for (a) $\psi = 10^{-5}$ and $\zeta=0.085$ (stationary instability) and 
        for (b) $\Psi=10^{-4}$ and $\zeta=0.3$ (oscillatory instability). The center of the convection rolls is shifted
        to $z_0 = -0.029$ and $z_0 = -0.005$, respectively. Here, $\lambda = \pi / q_c$.
}

\end{figure}

The symmetry breaking caused by the nonlinear ground state of the temperature and concentration field 
generates characteristic changes of the flow field. 
Let us first concentrate on the case $\psi > 0$. We found that for non vanishing values of $\zeta$ the 
center $z_0$ of the convection rolls is always shifted out of the center of the cell at $z=0$ towards the 
lower plate, i.e., $z_0 < 0$.
For a fixed value of $\psi>0$ the position $z_0$ decreases monotonically for rising values of $\zeta$
as long as the onset of convection is stationary and, besides this shift, no further qualitative deformation
of the convection rolls takes place. A typical flow field for this situation
is presented in Fig.~\ref{flowfield}(a) for $\psi = 10^{-5}$ and $\zeta = 0.085$, where we show contour
lines of the stream function $\Phi$.
This behavior changes, when $\zeta$ takes even larger values for which convection sets in
via a Hopf-bifurcation. In this paramter range we find inclined convection rolls, as exemplarily
presented in Fig.~\ref{flowfield}(b) for $\psi = 10^{-5}$ and $\zeta = 0.30$. 
Left travelling waves (LTW, with $\omega_c > 0$) are inclined to the left, 
right travelling waves (RTW, with $\omega_c < 0$) are inclined to the right.
Similarly inclined rolls appear also for
negative values of the separation ratio and finite values of $\zeta$, but in contrast to the
results for $\psi > 0$ the center is slightly shifted towards the upper boundary for $\psi < 0$. 

The dependence of the shift of the center of the convection rolls 
on the sign of the separation ratio $\psi$ can be understood as follows.
For  $\psi > 0$, the suspended particles are driven to the upper plate of the convection cell
and hence, the thermal conductivity rises from the lower to the upper plate. As the vertical
heat current $j_z = - \chi \partial_z T_{\, \rm{cond}}$ is independent of $z$ (compare
Eq.~\ref{heatcurrent}) it follows straight forward, that the gradient of the temperature
profile, which enhances the onset of convection, decreases from the lower to the upper plate.
Therefore, the fluid motion tends to concentrate in the region of the lower boundary, which
results into a shift of the center of the convection rolls to $z_0 < 0$. Obviously, the
situation is reversed for $\psi < 0$ and hence, the center of convection rolls is shifted upwards. 

%
\subsection{Free-slip, permeable boundary conditions\label{free-slip}} 
%
Let us now turn to free-slip, permeable boundary conditions.
It is well known that for free-slip, permeable boundary conditions~\eqref{bcsc2}
instead of no-slip, impermeable boundary conditions~\eqref{bcsc} the critical
wave number $q_c$ does not tend to zero for increasing values of $\psi$
\cite{Cross:88.2,Zimmermann:93.1}. In spite of this difference the major trends
are similar. For finite values of $\zeta$ the threshold is enhanced compared
to its value at $\zeta = 0$ in the range $\psi>0$ and lowered in the range $\psi < 0$.
As in the case of no-slip, impermeable boundary conditions we find a competition
between a stationary and an oscillatory onset of convection in the range $\psi > 0$,
which, however, reveals some differences, as described in this section.
\begin{figure}[ttt]
  \subfigure[\, Neutral curves at $\, \psi=10^{-4}$.]
  {
  \includegraphics[width=0.9\columnwidth]{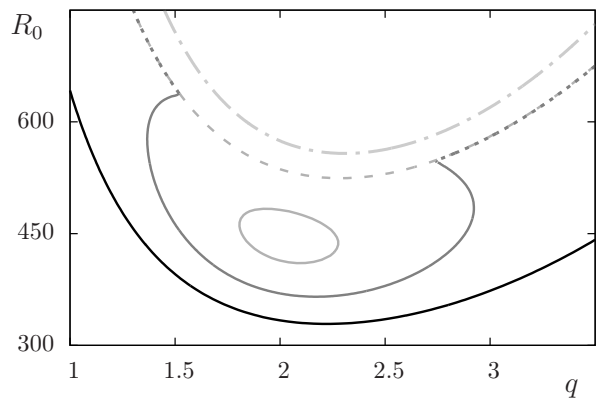}
  }
  \vspace{2mm}
  \subfigure[\, Neutral curves at $\, \psi=10^{-3}$.]
  {
  \includegraphics[width=0.9\columnwidth]{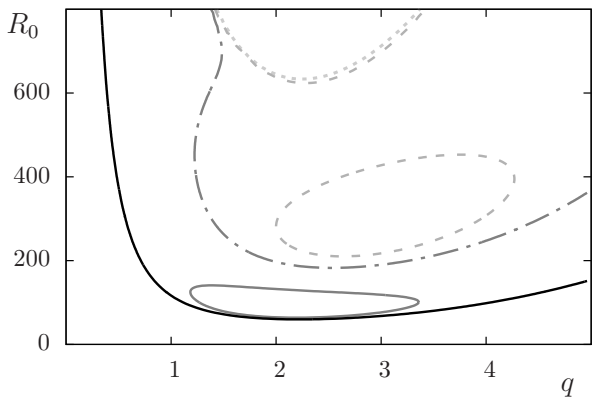}
  }
\vspace{-5mm}
  \caption{\label{neutstationary2}
           Neutral curves $R_0(q)$ for different values of the conductivity
           parameter $\zeta$ are shown in part (a) at $\psi=10^{-4}$
           with $\zeta=0$ (black solid line), $\zeta=0.08$ (dark-gray lines),
           $\zeta=0.095$ (gray lines) and  $\zeta = 0.50$ (light-gray dashed-dotted line)
           and in part (b) at $\psi=10^{-3}$ with $\zeta=0$ (black solid line),
           $\zeta=0.20$ (dark-gray lines), $\zeta=0.35$ (gray lines)  and  
           $\zeta=0.50$ (light-gray dotted line). Solid curves mark stationary
           instabilities and all ohter line types oscillatory ones. 
}
\end{figure}

\begin{figure}[ttt]
  \includegraphics[width=0.91\columnwidth]{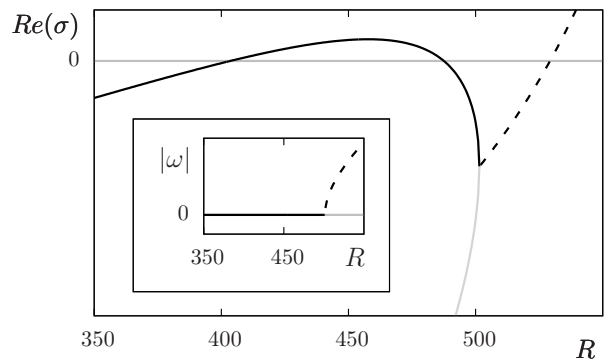}
\vspace{0mm}
  \caption{\label{sigma}
           The real parts of the two largest eigenvalues are shown as functions of 
           the Rayleigh number $R$ corresponding to the neutral curves in
           Fig.~\ref{neutstationary2}(a) for $\zeta=0.095$ and at $q=q_c\simeq 2.1$.
           The black solid line marks the stationary branch and has two zero crossings,
           which define the boundary of the closed curve in Fig.~\ref{neutstationary2}(a).
           At about $R\simeq 502$ this real eigenvalue merges with a second real
           eigenvalue (light-gray solid line) to a pair of complex conjugate eigenvalues
           (black dahed line). The appendant imaginary part $|\omega|$ is depicted in the inset.
}
\end{figure}

Fig.~\ref{neutstationary2} shows neutral curves for different values of $\zeta$ 
at $\psi = 10^{-4}$ and $\psi = 10^{-3}$. For smaller values of the separation ratio
the major effect of finite values of $\zeta$ is a shift towards higher Rayleigh numbers,
very similar to the situation depicted in Fig.~\ref{neutstationary1}(a).
For $\psi = 10^{-4}$ the neutral curves are additionally deformed and split into
a stationary branch and an oscillatory one at higher Rayleigh numbers as illustrated
in Fig.~\ref{neutstationary2}(a). Even though the situation is similar to the one shown
in Fig.~\ref{neutstatzeta_2} there are some interesting differences. For $\zeta=0.08$
the dark-gray solid curve merges with the dark-gray dotted curve without crossing, i.~e. two 
stationary eigenvalues merge and build a pair of complex conjugate ones similar as
presented in Fig.~\ref{sigma} for $\zeta=0.095$. The island at finite values of $q$
marked by the gray solid curve in Fig.~\ref{neutstationary2}(a) and corresponding
to $\zeta = 0.095$ is a further feature which differs from the case of no-slip boundary
conditions. The upper boundary of this curve marks a linear restabilization of the heat
conductive state, as indicated by the second sign change of the real part in Fig.~\ref{sigma}.
After the restabilization an even higher positioned Hopf bifurcation occurs, as described 
by the gray dashed line in Fig.~\ref{neutstationary2}(a), which in the ranges of small and 
large values of $q$ almost coincides with the oscillatory branch for $\zeta = 0.08$.
As this restabilization takes place
far beyond the first instability, where the nonlinear contributions to the equations of
motion determine the dynamics of the system, an answer to the question whether this
restabilization is of relevance for a real system can only be given by solving the
nonlinear equations of motion in that range.

For separation ratios larger than $\psi=10^{-4}$ even a linear restabilization of the
oscillatory bifurcation may occur. This is illustrated for $\psi=10^{-3}$ in
Fig.~\ref{neutstationary2}(b). For $\zeta=0.2$ we find a linear restabilization
of the stationary branch, as depicted by the dark-gray solid line, before an 
oscillatory bifurcation from the heat conductive state (dark-gray dashed-dotted line) takes place.
In contrast, for larger values of $\zeta$, e.~g., $\zeta=0.35$, the stationary branch
has already disappeared and the lowest instability is given by a Hopf bifurcation. Here,
this oscillatory branch shows a linear restabilization and a second Hopf bifurcation
arises at even higher values of $R$ as shown by the upper gray dashed curve in  
Fig.~\ref{neutstationary2}(b).

\begin{figure}[ttt]
  \subfigure[\, Critical Rayleigh number $R_c(\zeta)$.]
  {
  \includegraphics[width=0.9\columnwidth]{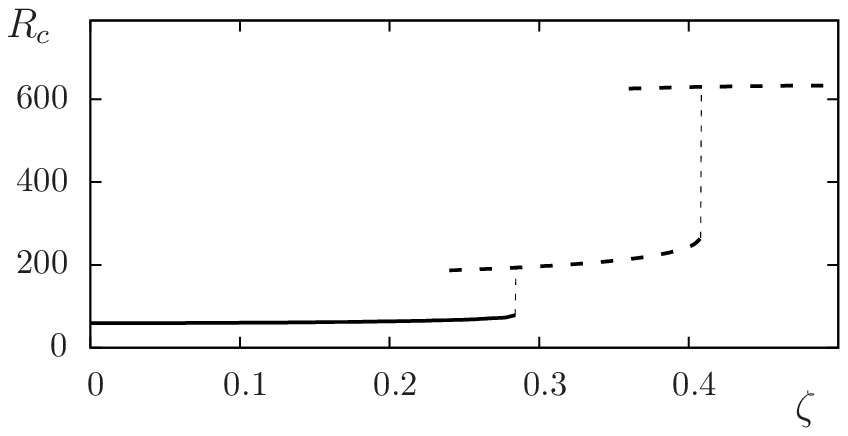}
  }
 \vspace{2mm}
  \subfigure[\, Critical wave number $q_c(\zeta)$.]
  {
  \includegraphics[width=0.9\columnwidth]{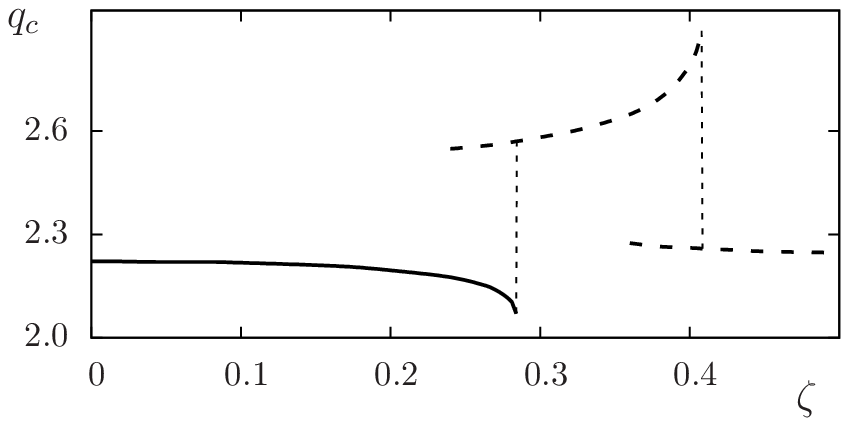}
  }
 \vspace{2mm}
  \subfigure[\, Critical frequency $\omega_c(\zeta)$.]
  {
  \includegraphics[width=0.9\columnwidth]{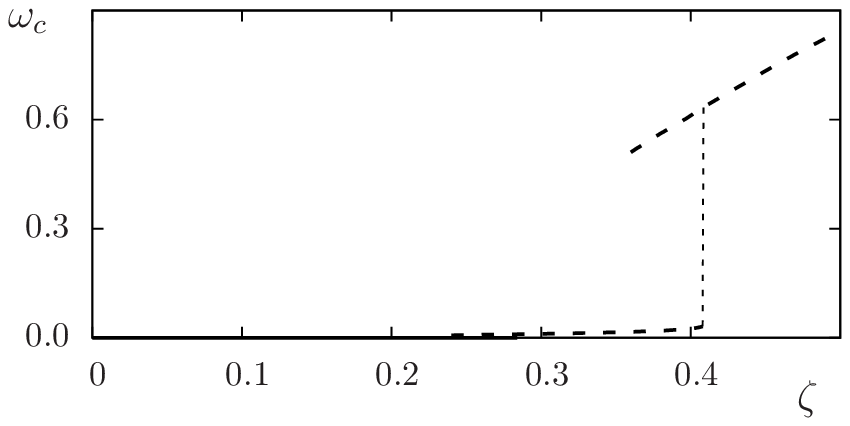}
  }
\vspace{-5mm}
\caption{\label{critvalues_fs}
         The critical Rayleigh number $R_c$ in (a), the critical wave number $q_c$ in (b),
         and the critical frequency $\omega_c$ in (c) are shown as functions of $\zeta$ 
         for $\psi = 10^{-3}$. Solid lines mark a stationary onset of convection
         and dashed lines an oscillatory one.
}
\end{figure}

These interesting bifurcation scenarios including two exchanges of instabilities 
are presented from a different perspective in Fig.~\ref{critvalues_fs}. Here, the 
critical values $R_c$, $q_c$, and $\omega_c$ are shown as functions of $\zeta$. 
In the first coexistence range the situation is similar to the case of rigid 
boundary conditions, i.~e., the end point of the solid line corresponds to the 
disappearance of an isola-like deformed stationary neutral curve. Similarly, in 
the second coexistence range the first oscillatory instability ceases to exist at 
the end point of the dashed line, leaving behind a second oscillatory instability 
at even higher values of the Rayleigh number. We mention that unlike the end points 
of the solid and first dashed line the starting points of both dashed lines are not 
related to the creation or aniquilation of isolas. We decided to display these lines 
also to the left of the transition points to stress the fact, that the oscillatory 
branches are already present before another branch vanishes, cf. Fig.~\ref{neutstationary2}.
The critical wave numbers
and frequencies corresponding to the three distinct instabilities differ considerably.   

\begin{figure}[ttt]
  \subfigure[\, Critical Rayleigh number $\, R_c(\psi)$.]
  {
  \includegraphics[width=0.9\columnwidth]{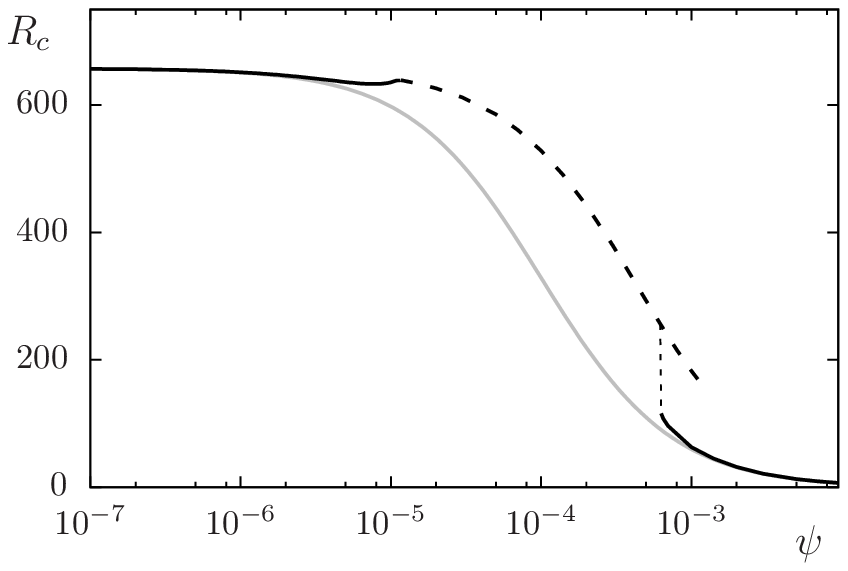}
  }
  \vspace{2mm}
  \subfigure[\, Critical wave number $\, q_c(\psi)$.]
  {
  \includegraphics[width=0.9\columnwidth]{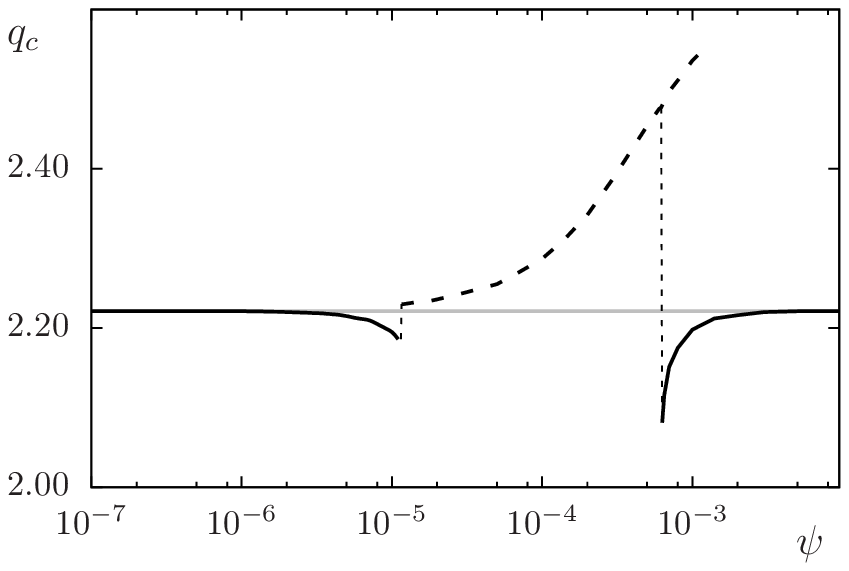}
  }
\vspace{-5mm}
  \caption{\label{RvsPsi_fs}
           Part (a) shows the critical Rayleigh number $R_c$ as a function
           of the separation ratio $\psi$ for $\zeta = 0$ (gray solid line)
           and $\zeta = 0.2$ (black lines), where in the latter case the solid lines
           mark the stationary branch and the dashed line the oscillatory
           one. Part (b) shows the corresponding critical wave number $q_c$ using
           the same line styles.
}
\end{figure}

A further illustration of the bifurcation scenario is given by Fig.~\ref{RvsPsi_fs},
where the critical Rayleigh number $R_c$ and the critical wave number $q_c$ are shown
as functions of the separation ratio $\psi$ for a conductivity parameter of $\zeta=0.2$.
Again, we find the most pronounced changes in comparison to $\zeta = 0$ in the range
$\psi \sim L = 10^{-4}$. When the two codimension-2 points marking the exchange of
instabilities are plotted as functions of $\zeta$, the resulting curve shows a similar
behavior as the one for no-slip boundary conditions in Fig.~\ref{Psivszeta}.

In accordance with the results for no-slip boundary conditions one obtains
for negative values of the separation ratio $\psi$ and finite values of $\zeta$
a reduction of the threshold compared to the case $\zeta=0$. The critical
wave number is independent from  $\zeta$ and takes the value $q_c = \pi / \sqrt{2}$
just like in the case of molecular binary-fluid mixtures.

For finite values of $\zeta$ the changes in the flow field at the onset of convection
with respect to the limiting case $\zeta = 0$ are qualitatively very similar to those discussed
for no-slip boundary conditions. While a positive separation ratio $\psi > 0$ lowers
the center of convection rolls, the opposite happens in the case $\psi < 0$. In addition,
an oscillatory onset leads to inclined rolls. As for free-slip
boundary conditions the velocity field takes finite values at both plates, the shift of the
center is larger than in the case of no-slip boundary conditions.

\section{Discussion and Conclusion\label{conclusions}} 
The onset of thermal convection in a colloidal suspension was investigated 
by means of a generalized continuum model for binary-fluid mixtures, which
has been extended beyond the Boussinesq approximation by taking into account
a linear dependence of the thermal conductivity on the local concentration of
colloidal particles.

We investigated colloidal suspensions where the thermal conductivity of the
suspension increases with rising values of the particle concentration. An
inhomogeneous particle concentration can be induced by temperature variations
via the Soret effect, which corresponds to finite values of the separation
ratio $\psi$. The concentration dependent thermal conductivity causes a nonlinear
variation of the temperature across the convection cell, in contrast to a linear
variation for a constant thermal conductivity. The strength of the spatial
variations of the heat conductivity is described by a dimensionless heat
conductivity parameter $\zeta$, as introduced in this work. It was found that
for finite values of $\zeta$ the vertical heat current through the convection
cell is reduced compared to the case of a constant heat conductivity.

For positive values of the separation ratio $\psi$ the suspended particles are driven
to the colder upper plate of the convection cell and for a constant heat conductivity,
i.~e., $\zeta=0$, a stationary bifurcation from the heat conductive state to convection
takes place. Spatial variations of the heat conduction, corresponding to $\zeta\not =0$,
lead to a shift of the critical Rayleigh number $R_c$ to larger values than obtained in
the case $\zeta=0$. Additionally, beyond some critical value $\zeta_c$ and in the parameter
range $\psi \sim L$ the onset of convection takes place via a Hopf bifurcation. This trend,
that a delay of the onset of convection leads to a Hopf bifurcation, is also met in the
range $\psi<0$, and well known from molecular binary-fluid mixtures.

Vertical variations of material properties, as discussed in this work for the thermal
conductivity, are considered to be important for modeling convection in the Earth's
mantle and, accordingly, a number of models including non-Boussinesq effects were
explored \cite{Bercovici:2009,Yuen:2001.1}. For modeling convection in systems with
spatially varying material parameters also two superimposed layers of 
immiscible \cite{Renardy:1985.1,Rasenat:1989.1,Andereck:1998.1} or even miscible fluids
\cite{Davaille:2002.1} are used. Like in our system with its spatially varying heat
conductivity, an oscillatory onset of convection was found in such two-layer systems 
for various parameters.

The range of positive $\psi$-values where the Hopf bifurcation takes place increases with rising values of
$\zeta$. At both boundaries of this range we find a codimension-2 bifurcation, where the
thresholds of the stationary and the oscillatory bifurcation coincide. Additionally, one has
a further codimension-2 bifurcation at $\psi \lesssim 0$, which is well known from earlier
investigations of molecular binary-fluid mixtures. However, for no-slip boundary conditions 
the two codimension-2 points occurring for $\psi>0$ are qualitatively different. Here, the
two neutral curves belonging to the two different bifurcations cross each other and therefore
three eigenvalues are close to be critical. In contrast to this, only two eigenvalues are
critical close to the codimension-2 bifurcation at $\psi \lesssim 0$. At the left boundary
of the $\psi$-range, where a Hopf bifurcation is preferred, one encounters usually small jumps
in the critical wave number, whereas large jumps occur at the right boundary of this range.

During our analysis we assumed $\gamma>0$ in Eq.~(\ref{eq:Einstein}), corresponding
to the situation that the thermal conductivity of the particles is higher than the one
of the base fluid. Hence, the local heat conductivity increases with a rising particle
concentration. However, we obtain for $\gamma <0$ exactly the same bifurcation scenarios
as described in this work for $\gamma>0$. This may be understood as follows. Independent
of the sign of $\gamma$, one obtains in the presence of thermophoresis layers of higher
and lower thermal conductivity accompanied by a nonlinear $z$-dependence of the temperature.
Changing the sign of $\gamma$ changes both, the variation of the heat conduction and the
temperature profile, but the vertical heat current is identical for both signs of $\gamma$,
as indicated by Eq.~(\ref{heatcurrent}). 

A further interesting question is how a spatially varying heat conductivity affects the ratio 
between the convective and conductive heat transfer (Nusselt number) beyond the  onset of 
convection, and how to understand the effects of boundary layers with enhanced or reduced heat 
conductivity.

\begin{acknowledgments}
Stimulating discussions with F.~H. Busse, M. Evonuk, G. Freund,
W. Pesch, I. Rehberg, and W. Sch\"opf are appreciated. This work
has been supported by the German science foundation through the
research unit FOR 608.
\end{acknowledgments}

\appendix*    
\section{Determination of the heat conductive state\label{app1}}
In the heat conductive state, i.~e. $\vec{v} = 0$, the temperature distribution
$T_{\, \rm{cond}}(z)$ and the particle distribution $N_{\, \rm{cond}}(z)$ are
determined by the two equations:
\begin{subequations}
\begin{align}
0 &= \partial_z \Big(\big(1+\gamma (N_{\, \rm{cond}}-N_0) \big) \partial_z T_{\, \rm{cond}}\Big)\,, \label{eq:ODE1}\\
0 &= \partial_z^2 N_{\, \rm{cond}}+\frac{k_T}{T_0}\,\partial_z^2T_{\, \rm{cond}}\,.                 \label{eq:ODE2}
\end{align}
\end{subequations}
For a vanishing mass current at the boundaries, cf. Eq.~\eqref{eq:boundary2},
a double integration of Eq.~\eqref{eq:ODE2} yields:
\begin{align}
 N_{\, \rm{cond}}=-\frac{k_T}{T_0}\,T_{\, \rm{cond}}+N_0+n_0\,.
\label{eq:NNN}
\end{align}
Now Eq.~\eqref{eq:ODE1} takes the form
\begin{align}
0  &= \partial_z^2 \left( (1+\gamma n_0) \, T_{\, \rm{cond}} - \frac{k_T}{2 T_0} T_{\, \rm{cond}}^2\right)  
\end{align}
and its integration gives
\begin{align}
T_{\, \rm{cond}}^2-\Gamma_0 \, T_{\, \rm{cond}} = \frac{1}{4}\left(C_1\,z+C_0\right) \, ,
\end{align}
with $\Gamma_0 = (1+\gamma n_0) T_0/ (\gamma k_T).$
Solving this quadratic equation the two unknown constants
$C_0$ and $C_1$ are determined by the two boundary 
conditions \eqref{eq:boundary3} for $T_{\, \rm{cond}}$ and one obtains:
\begin{align}
T_{\, \rm{cond}} = \Gamma_0 - \sqrt{
M_{+}^2\left(\frac{1}{2}-\frac{z}{d} \right) +
M_{-}^2\left(\frac{1}{2}+\frac{z}{d} \right)} \, ,
\end{align}
with $M_{\pm}=\Gamma_0+T_0\pm \delta T/2$. The last free constant
$n_0$ in Eq.~\eqref{eq:NNN} is determined by the definition of $N_0$:
\begin{align}
 N_0 = \frac{1}{d} \int\limits_{-d/2}^{d/2} N_{\, \rm{cond}}\,dz =
 N_0 + n_0 - \frac{k_T}{T_0 d}\int\limits_{-d/2}^{d/2} T_{\, \rm{cond}}\, dz\,.
\end{align}

\end{document}